\documentclass[10pt, journal]{IEEEtran}
\usepackage{graphicx}
\usepackage{caption}

\usepackage{amsmath,amsfonts}
\usepackage{algorithmic}
\usepackage{algorithm}
\usepackage{array}
\usepackage[caption=false,font=normalsize,labelfont=sf,textfont=sf]{subfig}
\usepackage{textcomp}
\usepackage{stfloats}
\usepackage{url}
\usepackage{verbatim}
\usepackage{graphicx}
\usepackage{cite} 
\usepackage{indentfirst}
\usepackage{graphicx}              
\usepackage{subcaption}            
\usepackage{graphicx}
\usepackage{bm}
\usepackage{amsmath}
\allowdisplaybreaks[4]
\usepackage{amssymb}
\usepackage{times}
\usepackage{mathtools}
\usepackage{psfrag}
\usepackage{cite}
\usepackage{lastpage}
\usepackage{fancyhdr}
\usepackage{color}
\usepackage{amsthm}
\usepackage{bigints}
\usepackage{multirow}
\usepackage{booktabs}
\usepackage{threeparttable}

\newtheorem{Theorem}{Theorem}

\theoremstyle{remark}

\newtheorem{theorem}[Theorem]{$\mathbf{Theorem}$}

\begin{document}
\title{Comparative Performance Analysis of Different Hybrid NOMA Schemes} 
\author{
Ning Wang, Chenyu Zhang, Yanshi Sun, ~\IEEEmembership{\itshape Member, IEEE}, \\ Minghui Min, ~\IEEEmembership{\itshape Senior Member, IEEE},
and Shiyin Li
\thanks{N. Wang, C. Zhang, M. Min and S. Li are with the School of Information and Control Engineering, China University of Mining and Technology, Xuzhou, 221116, China. (email: wangnsky@cumt.edu.cn, zcy6@cumt.edu.cn, minmh@cumt.edu.cn, lishiyin@cumt.edu.cn). 

Y. Sun is with the School of Computer Science and Information Engineering, Hefei University of Technology, Hefei, 230009, China. (email: sys@hfut.edu.cn). }
}
\maketitle

\begin{abstract}
Hybrid non-orthogonal multiple access (H-NOMA), which combines the advantages of pure NOMA and conventional OMA organically, has emerged as a highly promising multiple access technology for future wireless networks. Recent studies have proposed various H-NOMA systems by employing different successive interference cancellation (SIC) methods for the NOMA transmission phase. However, existing analyses typically assume a fixed channel gain order between paired users, despite the fact that channel coefficients follow random distribution, leading to their magnitude relationships inherently stochastic and time-varying. This paper analyzes the performance of three H-NOMA schemes under stochastic channel gain ordering: a) fixed order SIC (FSIC) aided H-NOMA scheme ; b) hybrid SIC with non-power adaptation (HSIC-NPA) aided H-NOMA scheme; c) hybrid SIC with power adaptation (HSIC-PA) aided H-NOMA scheme. Theoretical analysis derives closed-form expressions for the probability that H-NOMA schemes underperform conventional OMA. Asymptotic results in the high signal-to-noise ratio (SNR) regime are also developed. Simulation results validate our analysis and demonstrate the performance of H-NOMA schemes across different SNR scenarios, providing a theoretical foundation for the deployment of H-NOMA in next-generation wireless systems.
\end{abstract}

\begin{IEEEkeywords}
non-orthogonal multiple access (NOMA), hybrid NOMA (H-NOMA), hybrid successive interference cancellation (HSIC), channel order, power adaptation, energy efficiency.
\end{IEEEkeywords}

\section{Introduction}
With explosive growth in the number of connected devices and the continuous emergence of data-intensive applications, designing a multiple access technology that combines spectral efficiency and backward compatibility has become one of the core challenges in sixth-generation (6G) wireless communication systems. Orthogonal Multiple Access (OMA) schemes, which allocate mutually independent resource blocks to different users, have supported the evolution of cellular networks over the past few decades. However, in the face of the demands of massive connectivity and ultra-high data rates, their rigid resource partitioning approach has gradually become a bottleneck. Non-Orthogonal Multiple Access (NOMA), in contrast, allows users to share the same time-frequency resources through power-domain multiplexing, thus achieving significant capacity gains\cite{makki2020survey, you2021towards}. Recent studies have also shown that NOMA can integrate effectively with 6G enabling technologies, including reconfigurable intelligent surfaces (RIS) \cite{li2023achievable, zhu2020power}, integrated sensing and communications (ISAC) \cite{sun2024study}, fluid antenna systems (FAS) \cite{new2023fluid} and pinching antenna systems (PASS) \cite{Ding2025Pinching}.

Although NOMA offers significant theoretical gains, directly integrating it into OMA networks is impractical. The core mechanisms of traditional networks, such as resource allocation and interference suppression, are designed based on orthogonality. A sudden migration would require a large-scale protocol redesign and could lead to compatibility issues.  Hybrid NOMA (H-NOMA) was first proposed for mobile edge computing (MEC) scenarios \cite{Ding2019MEC}. By allowing users to access both dedicated orthogonal resources and shared NOMA resources, it retains backward compatibility while leveraging the benefits of NOMA, thus providing a viable solution to this challenge \cite{liu2021latency, wei2022energy, Ding2025HNOMAopt, Fang2025Rethinking}.

For the NOMA transmission phase in H-NOMA, successive interference cancellation (SIC) is essential for mitigating inter-user interference. Traditional fixed SIC (FSIC) methods predefine decoding orders based on channel state information (CSI) \cite{higuchi2013non, gao2017theoretical, Xia2018outage} or quality of service (QoS) requirements \cite{zhou2018state, Dhakal2019noma, ding2021new}, but suffer from irreducible error floors and suboptimal performance under dynamic channel conditions. In contrast, hybrid SIC (HSIC) dynamically adjusts decoding orders by jointly considering data rate requirements, transmission power, and instantaneous CSI, significantly improving transmission robustness \cite{ding2021new, ding2020unveiling1, ding2020unveiling2, Yang2023HSICpower}.

Existing studies on H-NOMA typically assume a fixed channel gain order between paired users \cite{sun2023hybrid, Sun2025HSICnPA, Wang2025HSICPA}(e.g., deterministic relationships between the channel coefficients of legacy and opportunistic users), while ignoring the stochastic and time-varying nature of wireless channels. In practice, channel coefficients are typically assumed to follow complex circularly-symmetric Gaussian distributions, making their magnitude relationships inherently random. The stochasticity of channel gains directly impacts SIC decoding efficiency and power allocation strategies, thereby influencing the overall performance of H-NOMA schemes. Thus, it is natural to ask the following question: how do different H-NOMA schemes perform when the channel gains are considered in a random order, which motivates this paper.

Building on previous research on H-NOMA schemes, this paper focuses on three representative schemes under stochastic channel gain ordering: (a) FSIC aided H-NOMA; (b) HSIC with non-power adaptation (HSIC-NPA) aided H-NOMA; and (c) HSIC with power adaptation (HSIC-PA) aided H-NOMA. The key contributions are as follows:
\begin{itemize}
\item A rigorous performance analysis of the three H-NOMA schemes is presented under stochastic channel gain ordering between paired users, with closed-form expressions derived for the probability that each scheme underperforms OMA.
\item The asymptotic analysis is derived in the high signal-to-noise ratio (SNR) regime, providing critical insights into the impact of the SIC order and power adaptation on performance.
\item Numerical validation of the theoretical results demonstrates that HSIC-PA outperforms the other schemes by eliminating performance floors in all scenarios.
\end{itemize}

This work aims to provide a theoretical foundation for the practical deployment of H-NOMA in next-generation wireless systems, where dynamic channel conditions and efficiency demands coexist.

\section{Signal Model}
Consider an uplink communication system consisting of one base station (BS) and $M$ users denoted as $U_i$, $1\le i\le M$. The channel gain between user $U_i$ and the BS is represented by $h_i$, and is assumed to be normalized small scale Rayleigh fading. Assume that the system initially operates under a time division multiple access (TDMA) legacy network, where each user is allocated a distinct time slot of duration $T$. For simplicity, assume that the $i$-th time slot is assigned to user $U_i$ in every frame. The transmission power of $U_i$ is denoted by $\rho_i$, then the achievable data rate of $U_i$ is given by $\log(1+\rho_i|h_i|^2)$. In this paper a slow time-varying channel is considered, where each user's channel gain is assumed to remain constant within a frame. Additionally, the background noise for all users is normalized to 1  

In H-NOMA schemes, users are classified into two types: legacy users and opportunistic users. 
Each legacy user is restricted to transmitting data only within its assigned time slot,  as in the legacy OMA network. In contrast, an opportunistic user can transmit data not only during its own time slot but also in the time slot of a legacy user by employing NOMA. Specifically, it is assumed that the opportunistic user $U_n$ is paired with a legacy user $U_m$, which means that $U_n$ can transmit its signal during both the $m$-th time slot (referred to as ``NOMA transmission'') and the $n$-th time slot (referred to as ``OMA transmission''). 
Recall that the channel coefficients $h_m$ and $h_n$ follow complex circularly symmetric Gaussian distributions, their magnitude relationship, $|h_m| \lessgtr |h_n|$, is inherently stochastic and may fluctuate over time. Consequently, it is essential to analyze the performance of H-NOMA based on the stochastic ordering relationship between $|h_m|$ and $|h_n|$.

To ensure a fair comparison, the power consumption of $U_n$ for the H-NOMA scheme is limited to $\beta\rho_n$ in NOMA and OMA time slots, where $0<\beta< 1/2$ denotes the power reduction coefficient as shown in \cite{Sun2025HSICnPA}. For the NOMA phase in the $m$-th slot, the outage probability for the transmission of $U_m$ should remain the same as that in the OMA scheme. This paper studies three different NOMA schemes: FSIC aided, HSIC-NPA aided, and HSIC-PA aided H-NOMA schemes, respectively. The achievable data rate for $U_n$ under different transmission schemes is introduced as in  \cite{Sun2025HSICnPA, Wang2025HSICPA}. 

If $U_n$ works in traditional OMA mode, it transmits data exclusively in the $n$-th time slot, with a transmission data rate of $R_n$:  
\begin{equation}\label{equ_Rn}
    R_n = \log  \left(1+\rho_n \left| h_n \right|^2 \right).
\end{equation}

If $U_n$ works in the H-NOMA mode, the achievable data rate in its assigned OMA time slot, i.e., the $n$-th time slot, is expressed as
\begin{equation}\label{equ_RnOMA}
    R^{\text{OMA}}_n = \log  \left (1+\beta\rho_n \left| h_n \right|^2 \right ).
\end{equation}
Meanwhile, in the NOMA time slot, i.e., the $m$-th time slot assigned for $U_m$, the achievable data rate of $U_n$ is different from each other since different NOMA schemes are applied, and will be analyzed as follows.
\subsection{FSIC aided H-NOMA scheme}
If $U_n$ works in the NOMA transmission slot of FSIC aided H-NOMA scheme, as shown in \cite{Sun2025HSICnPA}, its transmission rate should be limited to  
\begin{equation}\label{equ_RnTypeI}
\bar{R}^{\text{NOMA}}_n = \log \left(  1+  \frac{\beta \rho_n \left| h_n \right|^2  }{\rho_m \left| h_m \right|^2  +1} \right).
\end{equation}
\subsection{HSIC aided H-NOMA scheme}
If $U_n$ works in the NOMA transmission slot of HSIC-NPA aided H-NOMA scheme which is proposed in \cite{Sun2025HSICnPA}, the SIC strategies can be classified into two types based on the relationship between the received power of $U_n$ at BS and a threshold denoted by $\tau_m = \max\left \{ 0,|h_m|^2\alpha_m^{-1}-1\right \}$, which represents the maximum interference level that $U_m$ can still maintain the same outage performance as in the OMA scheme. Here $\epsilon_m = 2^{R_m}-1$, $\alpha_m = {\epsilon_m} /{\rho_m}$ and $R_m$ is the preset data rate of User $U_m$. 
The achievable rates for two different types are expressed as follows.
\subsubsection{Type I}
The received power of $U_n$ at the BS is less than or equal to the interference threshold, i.e.,  $\beta \rho_n \left| h_n \right|^2 \leq \tau_m$. In this scenario, the signal of $U_m$ is decoded in the first stage of SIC, and the achievable rate of $U_n$ is 
\begin{equation}
    \tilde{R}^I_n = \log \left(  1+  {\beta \rho_n \left| h_n \right|^2  } \right).
\end{equation}
\subsubsection{Type II}
The received power of $U_n$ at the BS is larger than the interference threshold, i.e., $\beta \rho_n \left| h_n \right|^2 > \tau_m$. In this scenario, the signal of $U_m$ is decoded at the first stage of SIC, and the achievable rate of $U_n$ is same as \eqref{equ_RnTypeI}
\begin{equation}
    \tilde{R}^{II}_n = \log \left(  1+  \frac{\beta \rho_n \left| h_n \right|^2  }{\rho_m \left| h_m \right|^2  +1} \right).
\end{equation}
\subsection{HSIC-PA aided H-NOMA scheme}
If $U_n$ works in the NOMA transmission slot of HSIC-PA aided H-NOMA scheme introduced in \cite{Wang2025HSICPA}. This scheme differs from HSIC-NPA primarily by introducing power adaptation in Type II, providing an opportunity to achieve higher achievable data rate. Similarly, the achievable rates of two different types are expressed as $\hat{R}^{I}_n$ and $\hat{R}^{II}_n$, respectively.
\subsubsection{Type I}
In this scenario, $\beta \rho_n \left| h_n \right|^2 \leq \tau_m$. The decoding process is the same as Type I of HSIC-NPA aided H-NOMA scheme
\begin{equation}
    \hat{R}^I_n = \log \left( 1 + {\beta \rho_n \left| h_n \right|^2 } \right).
\end{equation}
\subsubsection{Type II}
In this scenario, $\beta \rho_n \left| h_n \right|^2 > \tau_m$. Unlike Type II of HSIC-NPA aided H-NOMA scheme, there are two possible cases for data transmission and decoding order due to power adaptation is introduced here.
\begin{itemize}
\item Case 1: In this case, the signal of $U_n$ is decoded at the first stage of SIC, and $U_n$'s data rate is the same as the FSIC one: 
\begin{equation}
    \hat{R}^{II,1}_n =  \log \left(  1+  \frac{\beta \rho_n \left| h_n \right|^2 }{\rho_m \left| h_m \right|^2  +1} \right).
\end{equation}
\item Case 2: In this case, power adaptation is introduced by selecting a power adaptation factor $0< \gamma \leq 1$ such that $\gamma \beta \rho_n \left| h_n \right|^2 = \tau_m$. Consequently, the signal of $U_n$ can be decoded in the second SIC stage, yielding the following achievable rate of $U_n$ 
\begin{equation}
    \hat{R}^{II,2}_n =  \log \left( 1+  \tau_m \right).
\end{equation}
\end{itemize}

Accordingly, the achievable rate of  $U_n$ in Type II can be expressed as
\begin{align}
      \hat{R}^{II}_n  
      = \max \left\{\hat{R}^{II,1}_n,\hat{R}^{II,2}_n \right\} = \log \left(\hat{r}^{II}_n\right).      
\end{align}
where $\hat{r}^{II}_n = \max \left\{\left(1+  \tfrac{\beta \rho_n \left| h_n \right|^2}{\rho_m \left| h_m \right|^2  +1} \right), \left( 1+  \tau_m \right) \right\}$.

Therefore, the achievable data rate of $U_n$ in NOMA slot of the HSIC-PA aided H-NOMA scheme is
\begin{align}
\hat{R}^{\text{NOMA}}_{n}= 
\begin{dcases}
 \hat{R}^{I}_n, 
 &\beta \rho_{n}\left | h_{n} \right|^2 \le   \tau_m \\  
\hat{R}^{II}_n,   &\beta \rho_{n}\left | h_{n} \right |^2  > \tau_m.
\end{dcases}
\end{align}

The probabilities that the achievable rates of different H-NOMA schemes fail to outperform their OMA counterparts are given by:
\begin{align}
\label{equ_PnFSIC}\bar{P}_n &= \mathrm{Pr}\left(\bar{R}^{\text{NOMA}}_n  + R^{\text{OMA}}_n \leq R_n\right), \  {\text{for FSIC}}  \\
\label{equ_PnHSICnPA}\tilde{P}_n &= \mathrm{Pr}\left(\tilde{R}^{\text{NOMA}}_n  + R^{\text{OMA}}_n \leq R_n\right), \ {\text{for HSIC-PA}}\\ 
\label{equ_PnHSICPA}\hat{P}_n &= \mathrm{Pr}\left(\hat{R}^{\text{NOMA}}_n  + R^{\text{OMA}}_n \leq R_n\right), \ {\text{for HSIC-PA}}
\end{align}

Recall that the energy consumption for FSIC aided H-NOMA and HSIC-NPA aided H-NOMA is $2T\beta\rho_n$, while for the HSIC-PA aided H-NOMA is $(1+\gamma)T\beta\rho_n$, which are all smaller than the OMA counterpart, since $0<\beta <1/2$ and $0<\gamma\leq1$. 

\section{Performance Analysis for H-NOMA schemes with Random Channel Gains Order}
As described in the previous section, different H-NOMA schemes are introduced and their achievable data rates are derived. Assuming that $h_m$ and $h_n$ are independent and identically distributed (IID) circularly symmetric complex Gaussian random variables, this section rigorously analyzes the probability of these H-NOMA schemes underperforming the OMA baseline. Both closed-form exact results and asymptotic analysis of these probabilities are derived,  providing detailed performance comparisons.

\subsection{Analysis for Probability $\bar{P}_n$}
After simple mathematical calculations, the probability $\bar{P}_n$ for FSIC aided H-NOMA scheme which is shown in \eqref{equ_PnFSIC} can be expressed as:
\begin{align}
\begin{split}\label{equ_FSIC_PnD}
\bar{P}_n &= P\Bigg( \left(1 + \frac{\beta \rho_n \left| h_n \right|^2 }{\rho_m \left| h_m \right|^2 + 1} \right) 
                 \left(1 + \beta\rho_n \left| h_n \right|^2 \right) \\
          &\qquad  \leq \left(1 + \rho_n \left| h_n \right|^2 \right) \Bigg) \\
          &= P\left( |h_n|^{2} \leq \frac{(1-\beta)(1+\rho_m|h_m|^{2}) - \beta}{\beta^{2}\rho_n} \right)
\end{split}
\end{align}

Recall that the channel gain $h_m$ and $h_n$ are IID, 
the joint probability density function (pdf) of $|h_m|^{2}$ and $|h_n|^{2}$ is given by: 
\begin{align}\label{equ_pdf}   f_{|h_m|^{2},|h_n|^{2}}\left(x,y\right) = e^{-(x+y)}.
\end{align}

When the parameter $\lambda >0$,  the following definite integral holds:
\begin{align} 
    \int_{a}^{b} e^{-\lambda x} dx = \frac{1}{\lambda} \left( e^{-\lambda a} - e^{-\lambda b}   \right). 
\end{align}

Consequently, by applying the above joint pdf and definite integral functions, the exact value of $\bar{P}_n$ can be obtained:
\begin{align} \label{equ_FSIC_PnV}
\begin{split}    
\bar{P}_n &= \int_{0}^{+\infty}\int_{0}^{\frac{(1-\beta)(1+\rho_m|h_m|^{2})-\beta}{\beta^{2}
\rho_n}} e^{-(x+y)} dydx \\
&=1-\frac{1}{z_1}e^{-z_2}
\end{split}
\end{align}
where $z_1 = 1+\frac{(1-\beta)\rho_m}{\beta^2\rho_n}$, $z_2=\frac{1-2\beta}{\beta^2\rho_n}$. 

Using the Taylor series expansion, the two functions $f(x)$ and $g(x)$
can be approximated as follows: $f(x)=e^{-x}\approx 1$ and $g(x)=1-e^{-x}\approx x$ when $x\to 0$.
Hence, when $\rho_n \to \infty$, $\rho_m \to \infty$, and $\frac{\rho_n}{\rho_m}=\eta$ is a constant, $\bar{P}_n$ can be approximated as:
\begin{align}\label{equ_Approx_FSIC}
\bar{P}'_n \approx \frac{(1-\beta)}{\beta^2\eta+(1-\beta)}.
\end{align}

It can be seen that, for a given $0<\beta<1/2$, $\bar{P}'_n$ cannot approach 0. 

\subsection{Analysis for Probability $\tilde{P}_n$}
Recall that the achievable data rate of $U_n$ in the NOMA slot of the HSIC-NPA aided H-NOMA scheme takes two distinct values depending on the relationship between $|h_m|^2\alpha_m^{-1}$ and $\tau_m$. Therefore, the probability $\tilde{P}_n$ can be expressed as:
\begin{align} \label{equ_HSICnPA_PnD}   
\tilde{P}_n &= \underbrace{P\Big( \left(1+\beta \rho_n|h_n|^{2}\right)\left(1+\beta \rho_n|h_n|^{2}\right) \leqslant \left(1+\rho_n|h_n|^{2}\right), }_{\tilde{P}_{I}} \nonumber \\
&\qquad \underbrace{ \beta \rho_n |h_n|^{2} \leqslant \tau_m \Big)}_{\tilde{P}_{I}} \nonumber\\
&+ \underbrace{P\Bigg(\!\!\left(1\!+\!\frac{\beta \rho_n \left| h_n \right|^2}{\rho_m \left| h_m \right|^2  \!+\!1} \right)\!\! \left(1\!+\!\beta \rho_n|h_n|^2\right) \!\leqslant\! \left(1\!+\!\rho_n|h_n|^{2}\right), }_{\tilde{P}_{II}} \nonumber\\
&\qquad \underbrace{ \beta \rho_n |h_n|^{2} > \tau_m \Bigg)}_{\tilde{P}_{II}}.
\end{align}

After some simple mathematical calculations, the following theorem is obtained.

\begin{theorem}
The exact value of the probability $\tilde{P}_n$ is expressed as:
\begin{align}\label{equ_HSIC_NPA_PnV}
\!\!\tilde{P}_n \!=\!\!
    \begin{dcases}\!    V,&\!\!\!\!\epsilon_m \!>\!\frac{\beta}{1-\beta} \\
    \!V \!-\!\frac{1}{ z_4}e^{\frac{1}{\beta\rho_n}-z_4 z_5}\!+\!\frac{1}{z_1}e^{-(z_2+z_1 z_5)},& \!\!\!\!\epsilon_m \! \le \!\frac{\beta}{1-\beta}
    \end{dcases},   
\end{align}
where $z_1=1+\frac{(1-\beta)\rho_m}{\beta^2\rho_n}$, $z_2=\frac{1-2\beta}{\beta^2\rho_n}$, $z_3 = \frac{1-\beta}{\beta}\alpha_m$, $z_4 = 1+\frac{\alpha_m^{-1}}{\beta\rho_n}$, $z_5=\frac{1-\beta}{\beta\alpha_m^{-1}-(1-\beta)\rho_m}$, and
\begin{align}
    V = 1-e^{-(z_2+z_3)}+\frac{1}{ z_4}e^{\frac{1}{\beta\rho_n}}e^{-z_3 z_4}-\frac{1}{z_1}e^{- z_2}.
\end{align}

When $\rho_n \to \infty$, $\rho_m \to \infty$, and $\frac{\rho_n}{\rho_m}=\eta$ is a constant, the approximation of the probability $\tilde{P}_n$ is derived as:
\begin{align}\label{equ_Approx_HSIC_NPA}
\tilde{P}'_n\approx
\begin{dcases}
G(\epsilon_m)+E_r, &\epsilon_m>\frac{\beta}{1-\beta} \\
G(\bar{z}_5),&\epsilon_m\le\frac{\beta}{1-\beta}
\end{dcases}
\end{align}
where $E_r = \frac{1}{\bar{z}_4}-\frac{1}{\bar{z}_1}$, $\bar{z}_1=1+\frac{1-\beta}{\beta^2\eta}$, $\bar{z}_2=\frac{1-2\beta}{\beta^2\eta}$, $\bar{z}_3 = \frac{1-\beta}{\beta}\epsilon_m$, $\bar{z}_4=1+\frac{1}{\beta\epsilon_m\eta}$,  $\bar{z}_5=\frac{(1-\beta)\epsilon_m}{\beta-(1-\beta)\epsilon_m}$, and 
\begin{align}
G(x)=&\frac{1}{\rho_m^2}\frac{1}{2\beta\epsilon_m\eta}(\bar{z}_3^2- x^2)-\frac{1}{\rho_m^2}\frac{1}{\beta\rho_n}(\bar{z}_3 - x)\\
&+\frac{1}{\rho_m} \bar{z}_2(1+\frac{1}{\rho_m} x-\frac{1}{\rho_m} \bar{z}_3)+\frac{1}{\rho_m^2}\frac{(1-\beta) x^2}{2\beta^2\eta}. \nonumber
\end{align}

\begin{IEEEproof}
    Please refer to Appendix A.	
\end{IEEEproof}

\end{theorem}
From Theorem I, it can be observed that when $\rho_n \to \infty$, $\rho_m \to \infty$, and $\frac{\rho_n}{\rho_m}=\eta$ is a constant, if $\epsilon_m>\frac{\beta}{1-\beta}$, then $\tilde{P}_n$ cannot approach 0, but will instead approach $E_r$. In contrast, when $\epsilon_m\le\frac{\beta}{1-\beta}$, $\tilde{P}_n$ can approach 0.

\subsection{Analysis for Probability $\hat{P}_n$}
Similarly as \eqref{equ_HSICnPA_PnD}, the probability $\hat{P}_n$ is expressed as:
\begin{align}    
\hat{P}_n=&\underbrace{P\Big(\left(1+\beta \rho_n|h_n|^{2}\right)\left(1+\beta \rho_n|h_n|^{2}\right)\leqslant \left(1+\rho_n|h_n|^{2}\right),}_{\hat{P}_I}\nonumber\\
&\quad \underbrace{\beta \rho_n |h_n|^{2} \leqslant \tau_m\Big)}_{\hat{P}_I} \nonumber \\
&+\underbrace{P\Big(\hat{r}^{II}_n  \cdot \left(1+\beta \rho_n|h_n|^2\right) \leqslant  \left(1+\rho_n|h_n|^{2}\right),}_{\hat{P}_{II}}\nonumber\\
&\quad \underbrace{\beta \rho_n |h_n|^{2} > \tau_m\Big)}_{\hat{P}_{II}}.
\end{align}

Using the same approach as in Theorem 1 for the probability $\tilde{P}_n$, the following theorem for the probability $\hat{P}_n$ can be derived.
\begin{theorem}
The probability of $\hat{P}_n$ is expressed as:
\begin{align}    
\!\hat{P}_n \!=\!1\!-\! e^{-(z_2 + z_3)}\!+\!\frac{1}{z_1}e^{-z_2}(e^{-z_1x_2}\!-\!1) \!+\!\sum_{k=1}^{n_c}\!Q(u,\theta_k),
\end{align}
where $x_2= \frac{(\frac{\epsilon_m}{\beta}-1)+\sqrt{(\frac{\epsilon_m}{\beta}-1)^2+4\epsilon_m(\frac{1-\beta}{\beta})}}{2\rho_m}$,  
\begin{align}
\!Q(u,\theta_k)\!=\!\frac{u\pi}{n_c}\!\sqrt{1\!-\!\theta_k^2} \cdot e^{\left(\frac{1-\alpha_m^{-1}\left(u\theta_k+v\right)}{1-\beta\alpha_m^{-1}\left(u\theta_k+v\right)}\frac{1}{\rho_n}-\left(u\theta_k+v\right)\right)},
\end{align}
where $u=\frac{x_2- z_3}{2}$, $v=\frac{x_2+ z_3}{2}=u+z_3$, $\theta_{k}=\cos\left(\frac{2k-1}{2n_c}\pi\right)$,  and $n_c$ is the parameter for Gauss-Chebyshev approximation.

When $\rho_n \to \infty$, $\rho_m \to \infty$, and $\frac{\rho_n}{\rho_m}=\eta$ is a constant, $\hat{P}_n$ can be approximated as:
\begin{align}
\begin{split}\label{equ_Approx_HSIC_PA}  
\hat{P}'_n\approx&\frac{1}{\rho_m} \bar{z}_2+\frac{1}{\rho_m^2} \left(\bar{z}_2+\frac{1}{\beta\eta}\right)\left(\bar{x}_2-\bar{z}_3\right) \\
+&\frac{1}{\rho_m^2}\frac{(1-\beta)}{\beta^2\eta}\left(\frac{\bar{x}_2^2}{2}+\epsilon_m \ln\left(\frac{1}{\beta}-\frac{\bar{x}_2}{\epsilon_m}\right) \right),
\end{split}
\end{align}
where 
$\bar{x}_2=\frac{{(\frac{\epsilon_m}{\beta}-1)+\sqrt{(\frac{\epsilon_m}{\beta}-1)^2+4\epsilon_m(\frac{1-\beta}{\beta})}}}{2}$.

It can be seen that when $\rho_n \to \infty$, $\rho_m \to \infty$, and $\frac{\rho_n}{\rho_m}=\eta$ is a constant, the probability $\hat{P}_n$  approaches 0.

\begin{IEEEproof}
    Please refer to Appendix B.	
\end{IEEEproof}

\end{theorem}

\section{Numerical Results}
This section presents simulation results to validate the analysis and demonstrate the performance of different H-NOMA schemes. Recall that the noise power is normalized to $1$, SNR in this section is same as $\rho_n$, i.e., SNR = $\rho_n$. A detailed analysis of the impact of SNR and parameter $\eta$ on the probabilities of different H-NOMA schemes, i.e., $\bar{P_n}$, $\tilde{P_n}$, and $\hat{P_n}$, is provided. It is shown that the HSIC-NPA aided H-NOMA scheme significantly outperforms the pure OMA scheme and other H-NOMA schemes. 

\begin{figure}[htbp]
    \centering
    \includegraphics[width=1\linewidth]{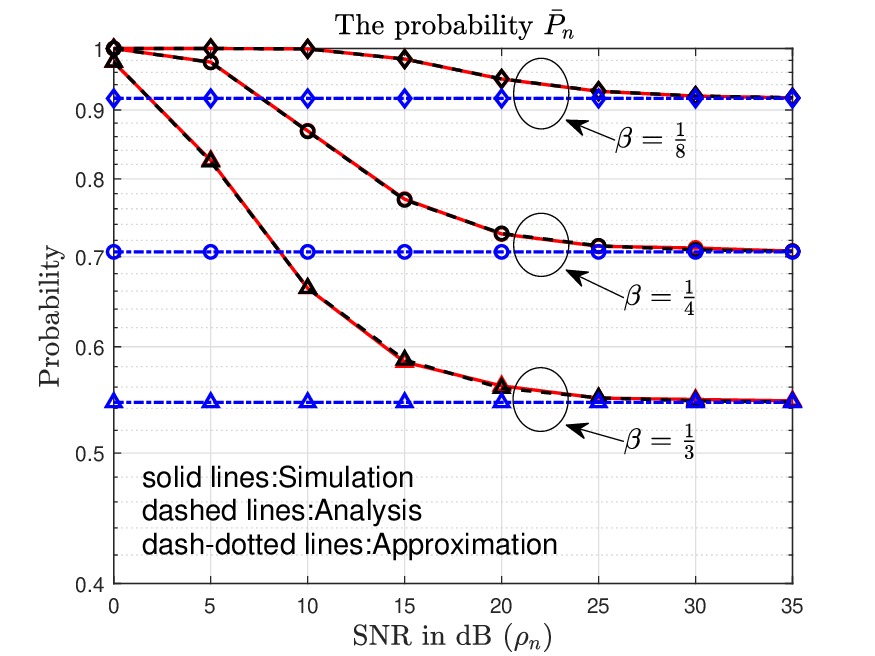}
    \caption{$\bar{P}_n$ versus SNR. $R_m=0.2$ bits per channel use (BPCU), $\eta=5$}
    \label{fig1_PnFSIC}
\end{figure}

\begin{figure}[htbp]
  \centering
  \setlength{\abovecaptionskip}{0em}
  \setlength{\belowcaptionskip}{0em}

  \subfloat[$\tilde{P}_n$ versus SNR. $R_m=1$ BPCU, $\eta=5$, $\epsilon_m>\frac{\beta}{1-\beta}$]
  {\includegraphics[width=1\linewidth]{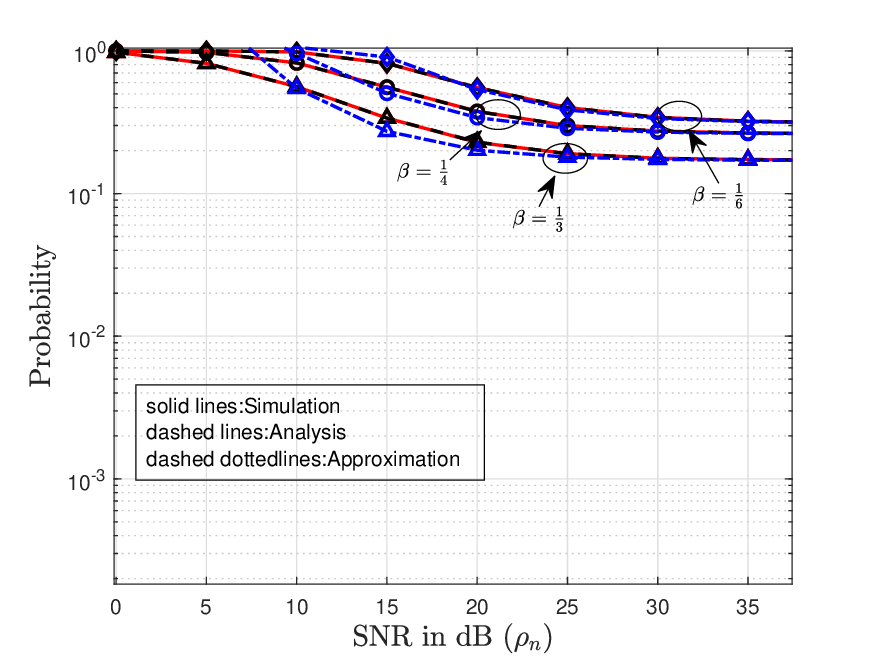}\label{fig3_PnHSICnPA1}}
  
  \subfloat[$\tilde{P}_n$ versus SNR. $R_m=0.1$ BPCU, $\eta=1$, $\epsilon_m\le\frac{\beta}{1-\beta}$]
  {\includegraphics[width=1\linewidth]{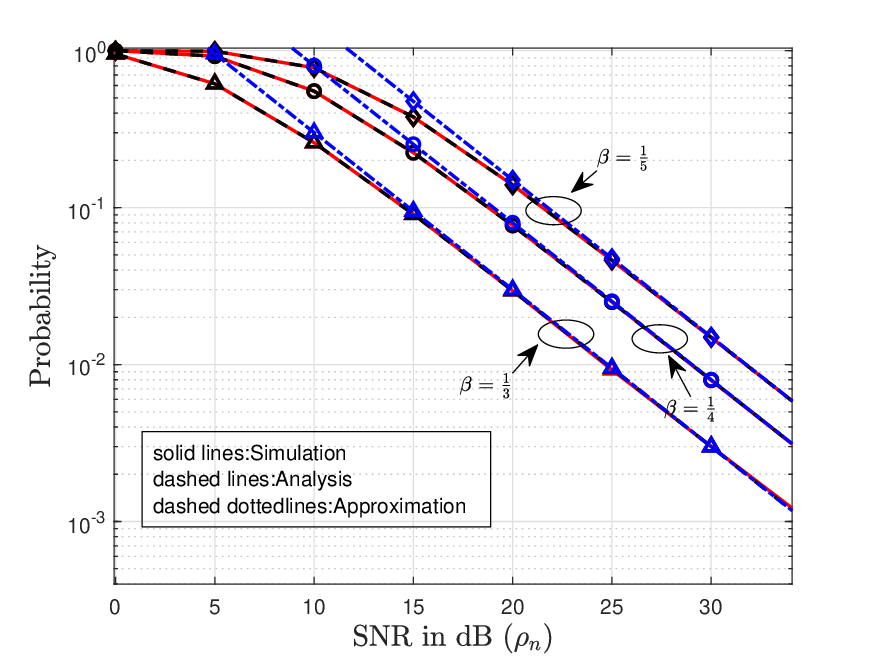}\label{fig3_PnHSICnPA2}}
  
  \vspace{0.5em}
  \caption{$\tilde{P}_n$ versus SNR for $\epsilon_m>\frac{\beta}{1-\beta}$ and $\epsilon_m\le\frac{\beta}{1-\beta}$}
  \label{fig2_PnHSICnPA}
\end{figure}

\begin{figure}[htbp]
    \centering
    \includegraphics[width=1\linewidth]{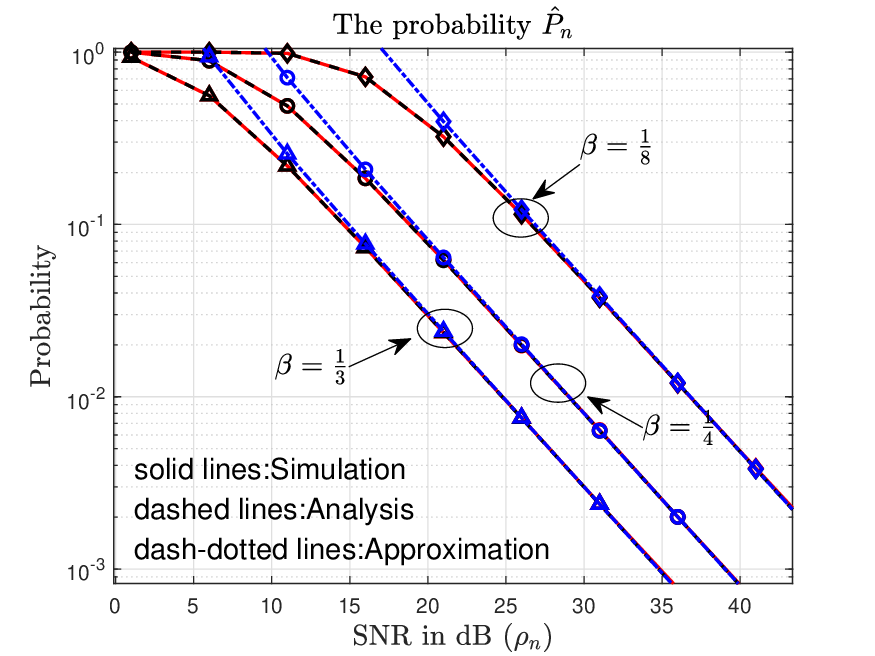}
    \caption{$\hat{P}_n$ versus SNR. $R_m=0.2$ BPCU, $\eta=5$}
    \label{fig3_PnHSICPA}
\end{figure}
Fig. \ref{fig1_PnFSIC}, \ref{fig2_PnHSICnPA}, and \ref{fig3_PnHSICPA} show the probability $\bar{P}_n$ for FSIC, $\tilde{P}_n$ for HSIC-NPA, and $\hat{P}_n$ for HSIC-PA aided H-NOMA schemes versus SNR, respectively.
As depicted in these figures, the analytical results and high-SNR approximations perfectly match the simulations, thereby verifying the accuracy of our developed analysis. 

As shown in Fig. \ref{fig2_PnHSICnPA}, the probability $\tilde{P}_n$ for $\epsilon_m>\frac{\beta}{1-\beta}$ and $\epsilon_m\le\frac{\beta}{1-\beta}$ is provided. It can be seen from the figures that, at high SNRs, when $\epsilon_m\le\frac{\beta}{1-\beta}$, $\tilde{P}_n$ can approach 0, whereas when $\epsilon_m>\frac{\beta}{1-\beta}$, $\tilde{P}_n$ exhibits a significant floor, which is generally consistent with the approximate results obtained earlier. In contrast, the probability $\hat{P}_n$ approaches 0 under all conditions. It can also be observed that, at high SNRs, the decreasing slopes of $\tilde{P}_n$ (when $\epsilon_m\le\frac{\beta}{1-\beta}$) and $\hat{P}_n$ are independent of the value $\beta$.
\begin{figure}[htbp]
	\centering
\subfloat[$R_m=1$ BPCU, $\frac{\rho_n}{\rho_m}=\eta=10$, $\beta=\frac{1}{4}$, $\epsilon_m>\frac{\beta}{1-\beta}$]
{\includegraphics[width=1\linewidth]{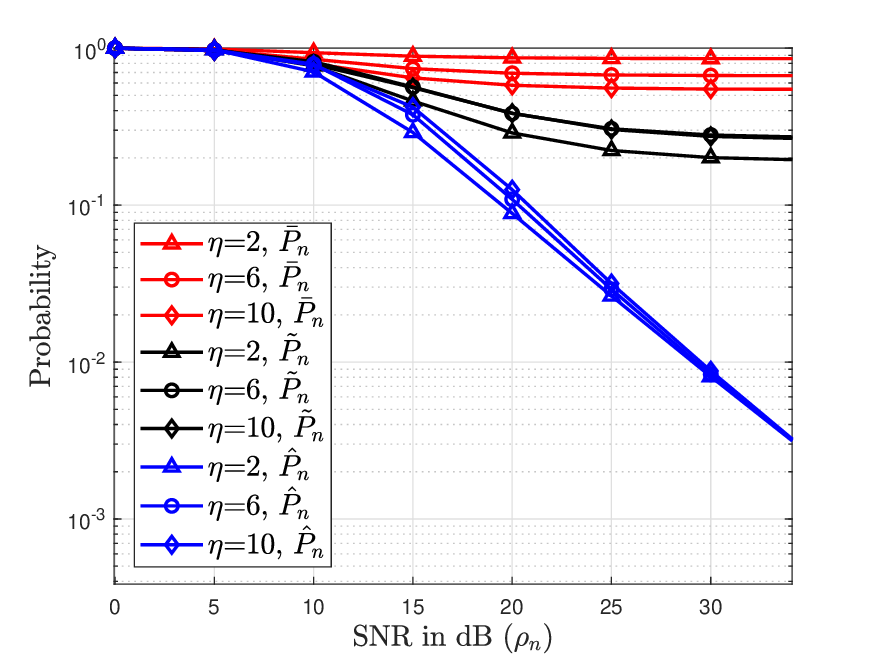}}
\label{fig4_a}
\subfloat[$R_m=0.1$ BPCU, $\frac{\rho_n}{\rho_m}=\eta=10$, $\beta=\frac{1}{3}$, $\epsilon_m\le\frac{\beta}{1-\beta}$]
{\includegraphics[width=1\linewidth]{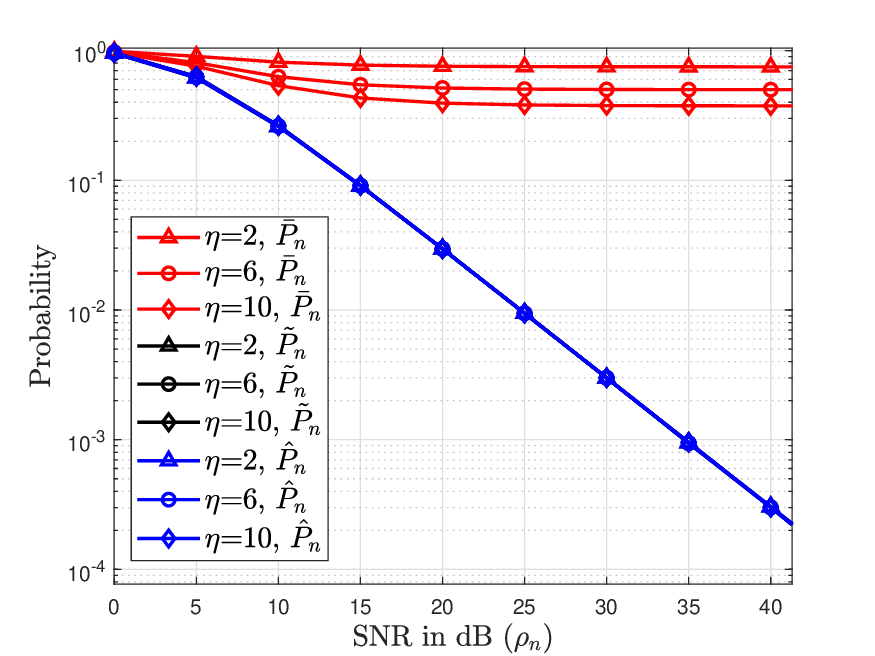}}
\label{fig4_b}
\vspace{0.5em}
\caption{Comparisons of FSIC, HSIC-NPA and HSIC-PA aided hybrid NOMA schemes in terms of $\bar{P}_n$, $\tilde{P}_n$ and $\hat{P}_n$, respectively, for $\epsilon_m>\frac{\beta}{1-\beta}$ and $\epsilon_m\le\frac{\beta}{1-\beta}$}
\label{fig4}
\end{figure}

Fig. \ref{fig4} compares the performance of FSIC, HSIC-NPA, and HSIC-PA aided H-NOMA schemes at high SNRs under different $\eta$. It can be seen from the figures that, at high SNRs, the slope of each probability is independent of $\eta$. Meanwhile, as depicted in Fig. \ref{fig4}(a), when $\epsilon_m>\frac{\beta}{1-\beta}$, only the probability $\hat{P}_n$ can approach 0 at high SNRs, while the other two probabilities $\bar{P}_n$ and $\tilde{P}_n$ have obvious severe floors. However, when $\epsilon_m\le\frac{\beta}{1-\beta}$, only the probability $\bar{P}_n$ has a severe floor. This is consistent with the previous results. Furthermore, it can be observed that HSIC-PA scheme exhibits superior performance compared to the other two schemes.

\section{Conclusion}
This paper evaluated the performance of three H-NOMA schemes (FSIC, HSIC-NPA, HSIC-PA) considering stochastic channel gain ordering. Closed-form expressions for the probability that each scheme underperforms OMA (i.e, \(\bar{P}_{n}\), \(\tilde{P}_{n}\), \(\hat{P}_{n}\)) were derived, along with their high-SNR asymptotic results. 
The analysis revealed that FSIC exhibits a non-zero performance floor, whereas HSIC-NPA only eliminates the floor under specific conditions. On the other hand, HSIC-PA, through the introduction of power adaptation, ensures that \(\hat{P}_{n}\) converges to zero in the high SNR regime across all scenarios, thereby outperforming the other schemes. The numerical simulations confirmed the theoretical findings, demonstrating the superior performance of HSIC-PA. This work established a theoretical foundation for the deployment of H-NOMA in next-generation wireless systems.

\appendices
\section{Proof for Theorem 1}
As shown in \textbf{Lemma 1} in \cite{Sun2025HSICnPA}, the probability $\tilde{P}_I$ can be further rewritten as: 
\begin{align}\label{equ_HSICnPA_Pn4}
    \tilde{P}_n&=\underbrace{P\left( |h_n|^2\le \Phi(|h_m|^2), \alpha_m<|h_m|^2< \frac{1-\beta}{\beta}\alpha_m\right)}_{\tilde{P}_{I,1}} \nonumber\\
            &+\underbrace{P\left(  |h_n|^2\le \frac{1-2\beta}{\beta^2\rho_n}, |h_m|^2> \frac{1-\beta}{\beta}\alpha_m \right)}_{\tilde{P}_{I,2}} \nonumber\\
            &+\underbrace{P\left( |h_n|^2\!\le\! \Psi(|h_m|^2), |h_n|^2\!>\!\Phi(|h_m|^2), |h_m|^2\!>\!\alpha_m \right)}_{\tilde{P}_{II,1}} \nonumber\\
            &+\underbrace{P\left(  |h_n|^2\le \Psi(|h_m|^2), |h_m|^2<\alpha_m  \right)}_{\tilde{P}_{II,2}},
\end{align}
where $\Psi(|h_m|^2)=\frac{(1-\beta)(1+\rho_m |h_m|^2)-\beta}{\beta^2\rho_n}$, and $\Phi(|h_m|^2)=\frac{|h_m|^2 \alpha_m^{-1}-1}{\beta\rho_n}$.

Since the probability $\tilde{P}_n$ consists of four probabilities, that is, $\tilde{P}_n=\tilde{P}_{I,1}+\tilde{P}_{I,2}+\tilde{P}_{II,1}+\tilde{P}_{II,2}$, their exact
values can be derived similarly to \eqref{equ_FSIC_PnV}. First, the probabilities $\tilde{P}_{I,1}$, $\tilde{P}_{I,2}$, and $\tilde{P}_{II,2}$ are directly expressed as follows:
\begin{align}\label{equ_NPA_PnV1}
\begin{split}
       &\tilde{P}_{I,1}=e^{-\alpha_m}-e^{- z_3}+\frac{1}{ z_4}e^{\frac{1}{\beta\rho_n}}\left(e^{- z_4 z_3}-e^{- z_4\alpha_m}\right),\\
       &\tilde{P}_{I,2}=\left( 1- e^{- z_2}\right)e^{-z_3},\\
       &\tilde{P}_{II,2}=1-e^{-\alpha_m}+\frac{1}{z_1}e^{- z_2}\left(e^{-z_1\alpha_m}-1\right),           
\end{split}
\end{align}
where $z_3 = \frac{1-\beta}{\beta}\alpha_m$, $z_4 = 1+\frac{\alpha_m^{-1}}{\beta\rho_n}$. 

Through a detailed comparison of $\Psi(|h_m|^2)$ and $\Phi(|h_m|^2)$, the probability $\tilde{P}_{II,1}$ can be categorized into the following results:
\begin{equation} \label{equ_tilde_P_II_1}
\tilde{P}_{II,1} = 
\begin{dcases}
P\!\Bigl( \Phi(|h_m|^2) < |h_n|^2 \le \Psi(|h_m|^2), \\
\quad |h_m|^2 > \alpha_m \Bigr), \ \ \ \ \ \quad
 \epsilon_m > \dfrac{\beta}{1-\beta} \\
P\!\Bigl( \Phi(|h_m|^2) < |h_n|^2 \le \Psi(|h_m|^2), \\
\quad \alpha_m < |h_m|^2 < z_5 \Bigr), 
 \ \epsilon_m \le \dfrac{\beta}{1-\beta} 
\end{dcases}
\end{equation}
where $z_5=\frac{1-\beta}{\beta\alpha_m^{-1}-(1-\beta)\rho_m}$.

Then, the value of $\tilde{P}_{II,1}$ can be expressed by applying the joint pdf in \eqref{equ_pdf} as:
\begin{equation}\label{equ_NPA_PnV2}
\tilde{P}_{II,1}=
    \begin{dcases}
    \frac{1}{z_4}e^{\frac{1}{\beta\rho_n}}e^{- z_4\alpha_m} 
    -\frac{1}{z_1}e^{- z_2}e^{- z_1\alpha_m},
    &\epsilon_m>\frac{\beta}{1-\beta}\\
    \frac{1}{z_4}e^{\frac{1}{\beta\rho_n}}\Bigl(e^{- z_4\alpha_m} 
    -e^{- z_4 z_5}\Bigr)\\
    \qquad -\frac{1}{z_1}e^{- z_2}\Bigl(e^{- z_1\alpha_m} 
    -e^{- z_1 z_5}\Bigr),
    &\epsilon_m\le\frac{\beta}{1-\beta}
    \end{dcases}
\end{equation}
Consequently, the exact value of the probability $\tilde{P}_n$ is obtained as \eqref{equ_HSIC_NPA_PnV}.

Furthermore, when $\rho_n \to \infty$, $\rho_m \to \infty$, and $\frac{\rho_n}{\rho_m}=\eta$ is a constant, each part of $\tilde{P}_n$, i.e., \eqref{equ_NPA_PnV1} and \eqref{equ_NPA_PnV2} can be approximated by using the Taylor series expansion as follows:
\begin{align}
\begin{split}
&\tilde{P}'_{I,1} \approx \frac{\alpha_m^{-1}}{2\beta\rho_n}( z_3^2-\alpha_m^2)-\frac{1}{\beta\rho_n}( z_3-\alpha_m)\\
&\tilde{P}'_{I,2} \approx  z_2(1- z_3) \\
&\tilde{P}'_{II,2} \approx  \frac{(1-\beta)\rho_m\alpha_m^2}{2\beta^2\rho_n}+ z_2\alpha_m 
\end{split}
\end{align}    

\begin{equation}
\tilde{P}_{II,1} \approx 
\begin{dcases}
\frac{1}{z_4} - \frac{1}{z_1}, \quad\quad\quad\quad\quad\quad\quad\quad \epsilon_m > \frac{\beta}{1-\beta} \\
\frac{(1-\beta)\rho_m(z_5^2 - \alpha_m^2)}{2\beta^2\rho_n} + z_2(z_5 - \alpha_m) \\
\quad \quad-\frac{z_5^2 - \alpha_m^2}{2\beta\rho_n\alpha_m} + \frac{z_5 - \alpha_m}{\beta\rho_n}, \epsilon_m \le \frac{\beta}{1-\beta}
\end{dcases}
\end{equation}

Therefore, the probability $\tilde{P}_n$ can be approximated as (\ref{equ_tilde_P_n_approx}).
\begin{figure*}[htbp!] 
\begin{equation} \label{equ_tilde_P_n_approx}
\tilde{P}_n \approx
\begin{dcases}
\frac{\alpha_m^{-1}}{2\beta\rho_n}( z_3^2-\alpha_m^2) - \frac{1}{\beta\rho_n}( z_3-\alpha_m) + z_2(\alpha_m + 1 - z_3)+ \frac{(1-\beta)\rho_m\alpha_m^2}{2\beta^2\rho_n} + \frac{1}{ z_4} - \frac{1}{ z_1},  &\epsilon_m > \frac{\beta}{1-\beta} \\
\frac{\alpha_m^{-1}}{2\beta\rho_n}( z_3^2 - z_5^2) - \frac{1}{\beta\rho_n}(z_3 - z_5) + z_2( z_5 + 1 - z_3) + \frac{(1-\beta)\rho_m z_5^2}{2\beta^2\rho_n}, &\epsilon_m \le \frac{\beta}{1-\beta}
\end{dcases}
\end{equation}
\end{figure*}
After performing some basic math operations,  $\tilde{P}_n$ can be further expressed as \eqref{equ_Approx_HSIC_NPA}.

\section{Proof for Theorem 2}
It is easy to see that $\hat{P}_I$ is the same as the probability $\tilde{P}_I$ shown in \eqref{equ_NPA_PnV1}:
\begin{align}
    \hat{P}_I = \tilde{P}_I = \tilde{P}_{I,1}+ \tilde{P}_{I,2}.
\end{align}

In this subsection, the probability $\hat{P}_{II}$ will be discussed in detail. 
Recall that when $|h_m|^{2} <\alpha_m$, $\tau_m=0$. Thus the probability $\hat{P}_{II}$ can be expressed as (\ref{equ_hat_P_II}). 

\begin{figure*}[htbp!] 
\begin{equation} \label{equ_hat_P_II}
\begin{split}
\hat{P}_{II} =& \underbrace{P\!\left( \left(1+\frac{\beta \rho_n|h_n|^{2}}{1+ \rho_m|h_m|^{2}}\right)\left(1+\beta\rho_n|h_n|^{2}\right) \leqslant \left(1+\rho_n|h_n|^{2}\right), |h_m|^{2} <\alpha_m \right)}_{\hat{P}_{II,1}} \\
&+ \underbrace{P\!\left( \hat{r}^{II}_n \cdot \left(1+\beta\rho_n|h_n|^{2}\right) \leqslant \left(1+\rho_n|h_n|^{2}\right), \beta\rho_n|h_n|^{2}>\tau_m, |h_m|^{2} >\alpha_m \right)}_{\hat{P}_{II,2}}.
\end{split}
\end{equation}
\end{figure*}
The probability $\hat{P}_{II,1}$ can be simplified as: 
\begin{align}
    \hat{P}_{II,1} = P\left(|h_n|^2 \leqslant \Psi(|h_m|^2), |h_m|^2<\alpha_m\right)
\end{align}

Meanwhile, since there are two different cases for $\hat{r}_n^{II}$, the probability $\hat{P}_{II,2}$ is divided into two components as (\ref{equ_hat_P_II_2}).
\begin{figure*}[htbp!] 
\begin{align} \label{equ_hat_P_II_2}
\hat{P}_{II,2} =& \underbrace{P\Bigg( \left(1 + \frac{\beta \rho_n \left| h_n \right|^2}{1+\rho_m \left| h_m \right|^2} \right) \left(1 + \beta \rho_n|h_n|^{2}\right) \leqslant \left(1+\rho_n |h_n|^{2}\right),\frac{\beta\rho_n |h_n|^{2}}{1+\rho_m |h_m|^2}>\tau_m,\ |h_n|^{2}>\Phi(|h_m|^2),\ |h_m|^{2}>\alpha_m \Bigg)}_{\hat{P}_{II,2,1}} \nonumber\\
 &+\underbrace{P\Bigg( \left(1+ \tau_m \right) \left(1+\beta \rho_n|h_n|^{2}\right) \leqslant \left(1+\rho_n |h_n|^{2}\right),\frac{\beta\rho_n |h_n|^{2}}{1+\rho_m |h_m|^2}<\tau_m,\ |h_n|^{2}>\Phi(|h_m|^2),\ |h_m|^{2}>\alpha_m \Bigg)}_{\hat{P}_{II,2,2}}.
\end{align}
\end{figure*}
There $\Omega(\left|h_{m}\right|^{2})=\frac{(\left|h_{m}\right|^{2}\alpha_{m}^{-1}-1)(1+\rho_{m}\left|h_{m}\right|^{2})}{\beta\rho_{n}}$. 
Hence, the probabilities $\hat{P}_{II,2,1}$ and $\hat{P}_{II,2,2}$ can be rewritten as:
\begin{align} 
\hat{P}_{II,2,1}&= P\Big(|h_n|^{2} <\Psi(|h_m|^2), |h_n|^{2}>\Omega(|h_{m}|^{2}),\notag \\
&\quad\quad\quad|h_n|^{2}>\Phi(|h_{m}|^{2}), |h_m|^{2}>\alpha_m \Big)  \\
\hat{P}_{II,2,2} &= P\bigg(|h_n|^{2}>\Theta(\left|h_{m}\right|^{2}),|h_n|^{2} < \Omega(\left|h_{m}\right|^{2}), \notag\\
&\quad\quad\quad|h_n|^{2}>\Phi(|h_{m}|^{2}), |h_m|^{2}>\alpha_m\bigg)
\end{align}
where $\Theta(\left|h_{m}\right|^{2})=\frac{(\left|h_{m}\right|^{2}\alpha_{m}^{-1}-1)}{(1-\beta|h_{m}|^{2}\alpha_{m}^{-1})\rho_{n}}$. 

It is easy to observe that $\frac{(|h_m|^{2}\alpha_m^{-1}-1)(1+\rho_m|h_m|^{2})}{\beta \rho_n} \geq \frac{|h_m|^{2}\alpha_m^{-1}-1}{\beta \rho_n}$, i.e., 
$\Omega(|h_{m}|^{2}) \geq \Phi(|h_{m}|^{2})$. Therefore,  
\begin{align}\label{equ_PB21}
\hat{P}_{II,2,1} = P\Big(&|h_n|^{2} <\Psi(|h_m|^2), |h_n|^{2}  >\Omega(|h_{m}|^{2}), \nonumber\\
&|h_m|^{2}>\alpha_m \Big).    
\end{align}

For the above probability to be meaningful, it must satisfy $\Omega(|h_{m}|^{2})<\Psi(|h_m|^2)$, 
which means 
\begin{align}\label{equ_hmfun}
|h_m|^{4}+(\frac{1}{\rho_m}-\frac{\alpha_m}{\beta})|h_m|^{2}-\frac{1-\beta}{\beta}\frac{\alpha_m}{\rho_m}<0.
\end{align}

Define $f(x) = x^2+(\frac{1}{\rho_m}-\frac{\alpha_m}{\beta})x-\frac{1-\beta}{\beta}\frac{\alpha_m}{\rho_m}, x=|h_m|^{2}\geq 0$. The two solutions of equation $f(x)=0$ are $x_1$ and $x_2$, expressed as 
\begin{align}
    \begin{dcases}
        x_1= \frac{(\frac{\epsilon_m}{\beta}-1)-\sqrt{(\frac{\epsilon_m}{\beta}-1)^2+4\epsilon_m(\frac{1-\beta}{\beta})}}{2\rho_m},\\
        x_2= \frac{(\frac{\epsilon_m}{\beta}-1)+\sqrt{(\frac{\epsilon_m}{\beta}-1)^2+4\epsilon_m(\frac{1-\beta}{\beta})}}{2\rho_m}.
    \end{dcases}
\end{align}

Since $x_1<0$, $x_2>0$, and it is easy to obtain $f(\alpha_m)<0$. Based on the properties of the roots of a quadratic equation, $x_1<\alpha_m <x_2$ holds. Then, \eqref{equ_PB21} should be rewritten as
\begin{align}
    \hat{P}_{II,2,1} = &P\Big( \Omega(|h_{m}|^{2}) < |h_n|^{2} <\Psi(|h_m|^2), \nonumber \\
    & \quad \alpha_m<|h_m|^2<x_2 \Big)
\end{align}

Since 
\begin{align}   |h_n|^{2}>\Theta(\left|h_{m}\right|^{2}) \Rightarrow  |h_m|^{2}<\frac{(1+\rho_n|h_n|^{2})\alpha_m}{1+\beta\rho_n|h_n|^{2}},
\end{align}
the following inequality holds due to $0<\beta<1/2$: 
\begin{align}
|h_m|^{2}<\frac{(\beta+\beta\rho_n|h_n|^{2})}{(1+\beta\rho_n|h_n|^{2})}\cdot \frac{\alpha_m}{\beta}<\frac{\alpha_m}{\beta},
\end{align}
which means that $(1-\beta|h_m|^{2}\alpha_m^{-1})>0$.

Recall that $|h_m|^{2}>\alpha_m$, therefore  
\begin{align}    
\Theta(\left|h_{m}\right|^{2}) = \frac{(|h_m|^{2}\alpha_m^{-1}-1)}{(1-\beta|h_m|^{2}\alpha_m^{-1})\rho_n} >0.
\end{align}

We then compare the magnitudes of variables $\Theta(\left|h_{m}\right|^{2})$ and $\Phi(\left|h_{m}\right|^{2})$, 
\begin{align}
    \begin{dcases}
        \Theta(\left|h_{m}\right|^{2}) \!<\! \Phi(\left|h_{m}\right|^{2}) \!\Rightarrow\! \alpha_m \!<\! |h_m|^2 \!<\! \frac{1-\beta}{\beta} \alpha_m\\
        \Theta(\left|h_{m}\right|^{2}) \!>\! \Phi(\left|h_{m}\right|^{2}) \!\Rightarrow\! \frac{1-\beta}{\beta} \alpha_m \!<\! |h_m|^2 \!<\! \frac{\alpha_m}{\beta} 
    \end{dcases}.
\end{align}

Hence, the probability $\hat{P}_{II,2,2}$ is divided into two distinct components:
\begin{align}
\begin{split}    
    \hat{P}_{II,2,2} =& \underbrace{P\Big(\Phi(\left|h_{m}\right|^{2}) < |h_n|^{2} 
    < \Omega(\left|h_{m}\right|^{2}),}_{\hat{P}_{II,2,2,1}}\\
    &\underbrace{\alpha_m<|h_m|^2<\frac{1-\beta}{\beta} \alpha_m \Big)}_{\hat{P}_{II,2,2,1}} \\
     + & \underbrace{P\Big(\Theta(\left|h_{m}\right|^{2})  < |h_n|^{2} < \Omega(\left|h_{m}\right|^{2}),}_{\hat{P}_{II,2,2,2}}\\
    &\underbrace{\frac{1-\beta}{\beta} \alpha_m<|h_m|^2<\frac{1}{\beta} \alpha_m \Big)}_{\hat{P}_{II,2,2,2}} 
\end{split}
\end{align}

It is easy to observe that the condition in the probability $\hat{P}_{II,2,2,1}$ always holds. And for probability $\hat{P}_{II,2,2,2}$, the following condition must be satisfied:
\begin{align}
&\Theta(\left|h_{m}\right|^{2}) < \Omega(\left|h_{m}\right|^{2})\notag \\
&\Rightarrow |h_m|^{4}+(\frac{1}{\rho_m}-\frac{\alpha_m}{\beta})|h_m|^{2}-\frac{1-\beta}{\beta}\frac{\alpha_m}{\rho_m}<0.
\end{align}

This condition is the same as \eqref{equ_hmfun}. It is easy to obtain $f\left(\frac{1-\beta}{\beta} \alpha_m\right)<0$ and $f\left(\frac{\alpha_m}{\beta}\right)>0$. Therefore, 
\begin{align}
    \hat{P}_{II,2,2,2} = &P\Big(\Theta(\left|h_{m}\right|^{2})  < |h_n|^{2} < \Omega(\left|h_{m}\right|^{2}),\notag\\
    &\quad \frac{1-\beta}{\beta} \alpha_m<|h_m|^2<x_2 \Big).
\end{align}

Based on the above analysis, the probability $\hat{P}_n$ consists of the following probabilities:
\begin{align}
    \hat{P}_n = \hat{P}_{I} + \hat{P}_{II,1} + \hat{P}_{II,2,1} + \hat{P}_{II,2,2,1} + \hat{P}_{II,2,2,2}
\end{align}

Consequently, each part of the probability can be obtained similarly to \eqref{equ_FSIC_PnV}:
\begin{align}
    &\hat{P}_{I} = -e^{-(z_2+z_3)} + e^{-\alpha_m}+\frac{1}{ z_4}e^{\frac{1}{\beta\rho_n}}(e^{- z_4 z_3}-e^{- z_4\alpha_m}),\notag\\
    &\hat{P}_{B,1} =1-e^{-\alpha_m}+\frac{1}{z_1}e^{- z_2}(e^{-z_1\alpha_m}-1),\notag  \\
    &\hat{P}_{B,2,1} ={G(x_2,\alpha_m)}-e^{- z_2}\frac{1}{ z_1}\left[e^{- z_1\alpha_m}-e^{- z_1x_2}\right], \\
    &\hat{P}_{B,2,2,1} =\frac{1}{ z_4}e^{\frac{1}{\beta\rho_n}}(e^{- z_4\alpha_m}-e^{- z_4 z_3})-G(z_3,\alpha_m),  \notag \\
    &\hat{P}_{B,2,2,2} =\sum_{k=1}^{n_c}Q(u,\theta_k)-{G(x_2,z_3)},\notag
\end{align}
where $z_1 = 1+\frac{(1-\beta)\rho_m}{\beta^2\rho_n}$, $z_2=\frac{1-2\beta}{\beta^2\rho_n}$, $z_3 = \frac{1-\beta}{\beta}\alpha_m$,  $z_4 = 1+\frac{\alpha_m^{-1}}{\beta\rho_n}$, 
$A=\frac{\alpha_m^{-1}\rho_m}{\beta\rho_n}$, $B=1+\frac{\alpha_m^{-1}-\rho_m}{\beta\rho_n}$, 
\begin{align}
G(x,y)=&e^{(\frac{1}{\beta\rho_n} +\frac{B^2}{4A})} \frac{e^{}\sqrt{\pi}}{2\sqrt{A}}\bigg[erf\left(\sqrt{A}(x+\frac{B}{2A})\right) \\ &-erf\left(\sqrt{A}(y+\frac{B}{2A})\right)\bigg] \nonumber\\
    Q(u,\theta_k)=&\frac{u\pi}{n_c}\sqrt{1-\theta_k^2} \\ &\times \exp{\left(\frac{1-\alpha_m^{-1}\left(u\theta_k+v\right)}{1-\beta\alpha_m^{-1}\left(u\theta_k+v\right))}\frac{1}{\rho_n}-\left(u\theta_k+v\right)\right)}, \nonumber
\end{align}
where $u=\frac{x_2- z_3}{2}$, $v=\frac{x_2+ z_3}{2}=u+z_3$, $\theta_{k}=\cos\left(\frac{2k-1}{2n_c}\pi\right)$, and $n_c$ is the parameter for Gauss-Chebyshev approximation as shown below. 
\begin{align} \label{equ_GaussCheb}
\begin{split}
    \int_{a}^{b}f(x)dx=&\sum_{i=1}^{n_c}\frac{\pi}{n_c}(\frac{a}{2}-\frac{b}{2})\\
    &\times f[(\frac{a}{2}+\frac{b}{2})+(\frac{a}{2}-\frac{b}{2})t_i]\sqrt{1-t_i^{2}}.    
\end{split}
\end{align}

Finally, the probability $\hat{P}_n$ can be obtained as:
\begin{align}    
\hat{P} _n =&1- e^{-(z_2 + z_3)}+\frac{1}{z_1}e^{- z_2}(e^{-z_1x_2}-1) \nonumber\\ &+\sum_{k=1}^{n_c}Q(u,\theta_k)
\end{align}

When $\rho_n \to \infty$, $\rho_m \to \infty$, and $\frac{\rho_n}{\rho_m}=\eta$ is a constant, each part of $\hat{P}_n$ can be approximated by applying Taylor series as follows:
\begin{align} 
    &\hat{P}_{I}\approx  z_2(1- z_3) + \frac{\alpha_m^{-1}}{2\beta\rho_n}( z_3^2-\alpha_m^2)-\frac{1}{\beta\rho_n}( z_3-\alpha_m), \nonumber\\
    &\hat{P}_{B,1}\approx  \frac{(1-\beta)\rho_m\alpha_m^2}{2\beta^2\rho_n}+ z_2\alpha_m, \nonumber\\
    &\hat{P}_{B,2,1}\approx  \frac{(1-\beta)\rho_m(x_2^2-\alpha_m^2)}{2\beta^2\rho_n}+ z_2(x_2-\alpha_m)+\frac{x_2-\alpha_m}{\beta\rho_n}\nonumber \\
    & \qquad\qquad -\frac{\alpha_m^{-1}\rho_m(x_2^3-\alpha_m^3)}{3\beta\rho_n}+\frac{(\rho_m-\alpha_m^{-1})(x_2^2-\alpha_m^2)}{2\beta\rho_n},\nonumber \\
    &\hat{P}_{B,2,2,1}\approx  \frac{\alpha_m^{-1}\rho_m( z_3^3-\alpha_m^3)}{3\beta\rho_n}-\frac{\rho_m( z_3^2-\alpha_m^2)}{2\beta\rho_n},  \\
    &\hat{P}_{B,2,2,2}\approx \frac{\alpha_m^{-1}\rho_m(x_2^3- z_3^3)}{3\beta\rho_n}+\frac{(\alpha_m^{-1}-\rho_m)(x_2^2- z_3^2)}{2\beta\rho_n} \nonumber\\
    & \qquad\qquad+\frac{(1-\beta)}{\beta^2\rho_n\alpha_m^{-1}}\ln\bigg(\frac{1-\beta\alpha_m^{-1}x_2}{1-\beta\alpha_m^{-1} z_3}\bigg),\nonumber
\end{align}
   
Therefore, $\hat{P}_n$ can be approximated as:
\begin{align}
\label{approximation1}    
\hat{P}'_n\approx& z_2(1- z_3)+\frac{(1-\beta)\rho_mx_2^2}{2\beta^2\rho_n}+ z_2x_2 \\
&+\frac{1}{\beta\rho_n}(x_2- z_3)+\frac{(1-\beta)}{\beta^2\rho_n\alpha_m^{-1}}\ln\bigg(\frac{1-\beta\alpha_m^{-1}x_2}{1-\beta\alpha_m^{-1} z_3}\bigg) \nonumber\\
=&\frac{1}{\rho_m} \bar{z}_2-\frac{1}{\rho_m^2} \bar{z}_3 \bar{z}_2+\frac{1}{\rho_m^2}\frac{(1-\beta)\bar{x}_2^2}{2\beta^2\eta}+\frac{1}{\rho_m^2} \bar{z}_2\bar{x}_2 \nonumber\\
&+\frac{1}{\rho_m^2}\frac{1}{\beta\eta}(\bar{x}_2- \bar{z}_3)+\frac{1}{\rho_m^2}\frac{(1-\beta)\epsilon_m}{\beta^2\eta}\ln\bigg(\frac{1-\frac{\beta\bar{x}_2}{\epsilon_m}}{\beta}\bigg)\nonumber 
\end{align}

After performing some basic math operations, $\hat{P}'_n$ can be further expressed as \eqref{equ_Approx_HSIC_PA}.

\bibliographystyle{IEEEtran}
\bibliography{IEEEabrv, ref}
\end{document}